\documentclass{iopjournal}
\usepackage[utf8]{inputenc}
\usepackage{graphicx}
\usepackage{lmodern}

\newcommand{\ddt}[1]{\frac{\mathrm{d}#1}{\mathrm{d}t}}
\newcommand{\Tp}{T_\mathrm{p}}

\NewDocumentCommand{\ioni}{mmm}{%
  \ifnum#3=1
    $^{#1}\mathrm{#2}^{+}$%
  \else
    $^{#1}\mathrm{#2}^{#3+}$%
  \fi
}
\NewDocumentCommand{\ion}{mm}{%
  \ifnum#2=1
    $\mathrm{#1}^{+}$%
  \else
    $\mathrm{#1}^{#2+}$%
  \fi
}

\begin{document}

\articletype{Paper} 

\title{Deposition rate and energy to substrate in chopped and standard HiPIMS: identifying optimal pulse parameters}

\author{M. Farahani$^1$\orcid{0009-0009-9444-3796}, J. Čapek$^1$\orcid{0000-0002-3267-7708} and T. Kozák$^{1,*}$\orcid{0000-0002-5046-7253}}

\affil{$^1$Department of Physics and NTIS -- European Centre of Excellence, University of West Bohemia in Pilsen, Univerzitní 8, 301 00 Plzeň, Czech Republic}

\email{kozakt@fav.zcu.cz}


\begin{abstract}
High-Power Impulse Magnetron Sputtering (HiPIMS) delivers higher ionisation of the target material at the expense of lower deposition rates than conventional magnetron sputtering. A fine optimisation of the deposition conditions is crucial for specific applications. Chopped or multi-pulse HiPIMS (segmenting pulses into shorter micropulses) has been proposed to enhance the degree of ionisation of the film-forming flux fraction while maintaining the deposition rate. This study investigates how micropulse length and the delay time between segments influence energy flux, deposition rate, and the ionised flux fraction in both chopped and standard HiPIMS. These quantities are evaluated by passive thermal probe, biasable QCM and mass spectrometer measurements at the substrate position. Deposition-averaged and pulse-averaged power is kept constant for all conditions to facilitate meaningful comparison.

Results indicate that chopping the HiPIMS pulse consistently leads to higher energy flux and total deposition rate compared to standard HiPIMS at the same total pulse length, primarily due to increased ion flux. In chopped HiPIMS, increasing the delay time between micropulses beyond $30\,\mu$s simultaneously improved the deposition rate and the ionised flux fraction. In comparison, standard HiPIMS, operated at higher frequencies and shorter pulse lengths, demonstrated similar performance with even higher total energy fluxes, but this can be attributed to a higher fraction of process-gas ions in the flux to the substrate, rather than to an increased ionised flux fraction of the film-forming species. 
\end{abstract}

\section{Introduction}

High-Power Impulse Magnetron Sputtering (HiPIMS) is an advanced sputtering technique that has gained considerable attention for its ability to generate dense plasma by applying high power in short pulses at a low duty cycle \cite{Lundin2019}. This approach results in a higher degree of ionisation of the sputtered material compared to conventional direct-current magnetron sputtering (DCMS). The resulting benefits include the deposition of dense, well-adhered films with the desired microstructure and improved conformity on complex substrate geometries \cite{Batkova2020, Kumar2020, Zeman2017}. 

Despite its advantages, HiPIMS typically exhibits a lower deposition rate than DCMS at the same average power. The decrease in the deposition rate is mainly attributed to the return of ionised sputtered species to the target and to other effects related to the depletion of the working gas \cite{Anders2010}. There is a trade-off between the deposition rate and the fraction of target-material ions in the total flux of target material to the growing film (the so-called ionised flux fraction, IFF). Ideally, one would want to have a high degree of ionisation of film-forming species in the dense plasma in front of the target (high ionisation probability) while allowing the ions to escape towards the substrate (low ion return probability) \cite{Vlcek2010, Anders2012a}.  This has become a focal point for innovation in reimagining the pulse architecture of HiPIMS to optimise the balance between the deposition rate and IFF. 

One such approach, which has been studied extensively, involves chopping a standard HiPIMS (s-HiPIMS) pulse into multiple micropulses \cite{Hnilica2026, Hnilica2025, Hnilica2023, Fekete2017, Antonin2015, Barker2013, Soucek2021, Huo2021}. This approach, called chopped HiPIMS (c-HiPIMS) or multipulse HiPIMS, has shown potential to increase the deposition rate while maintaining the IFF, or to enhance the IFF while maintaining the deposition rate. As reported in our earlier work on multi-pulse bipolar HiPIMS \cite{Farahani2024}, unipolar c-HiPIMS yields higher energy fluxes than s-HiPIMS at the same average power and total pulse length. These findings have mainly been explained by the lower effective ion return probability (to the target) due to the frequent off-times between short pulses when the strong electric field directed toward the target disappears or partial recovery of the process gas density after rarefaction during the pulse \cite{Rudolph2020, Hnilica2023, Barker2014}. 

The parameter space in HiPIMS is quite broad, and published studies often employ different operating conditions and pulse chopping strategies. As a result, their findings are not easily comparable or generalizable. While the average power (in the period) is typically held constant, the total pulse length or the average pulse power (averaged over the total on-time) is often varied. When the average pulse power (or similarly the peak discharge current) varies considerably, the plasma density and, consequently, the degree of ionisation of metal atoms in the plasma also vary considerably. In that case, it is difficult to separate the effects of changing ionisation probability and ion return probability. For example, in recent work of Hnilica et al.\cite{Hnilica2025, Hnilica2026}, increasing the number of pulses while decreasing the average pulse power results in a predominant decrease in the IFF due to the lower ionisation probability. When the authors held the pulse power constant, the IFF decreases slightly with increasing number of pulses, which seems to contradict our measurements of energy flux \cite{Farahani2024}.

Due to these differences in operating conditions, it remains unclear, in our view, whether the benefits of c-HiPIMS stem from the specific pulse architecture or simply from operating in a shorter-pulse regime. Therefore, in this study, our objective is to deepen the understanding of c-HiPIMS by systematically investigating how variations in the length of the micropulses, the delay time, and the magnetic field strength influence the deposition characteristics (energy flux, deposition rate, and ionised flux fraction of target material species) at the substrate position. To ensure meaningful comparisons and maintain a consistent ionisation degree across all experiments, both the average power and (average) pulse power were kept constant for all configurations, which is not commonly emphasised in earlier investigations. In parallel, variations in pulse length and repetition frequency in s-HiPIMS (without pulse chopping) are also explored to compare the effectiveness of the standard and chopped approaches. A combination of diagnostic tools was employed: a passive thermal probe and a QCM ion flux meter to measure the total energy and the sputtered metal atom and ion fluxes to the substrate, respectively, and a mass spectrometer to separately quantify singly and doubly ionised titanium and argon species at the same location as the probes. These complementary measurements provide detailed insight into the plasma characteristics during chopped and standard HiPIMS operation. This understanding is critical for the optimal application of HiPIMS in film deposition. 

\section{Experimental details and methodology}

\begin{figure} 
    \centering
    \includegraphics[width=85mm]{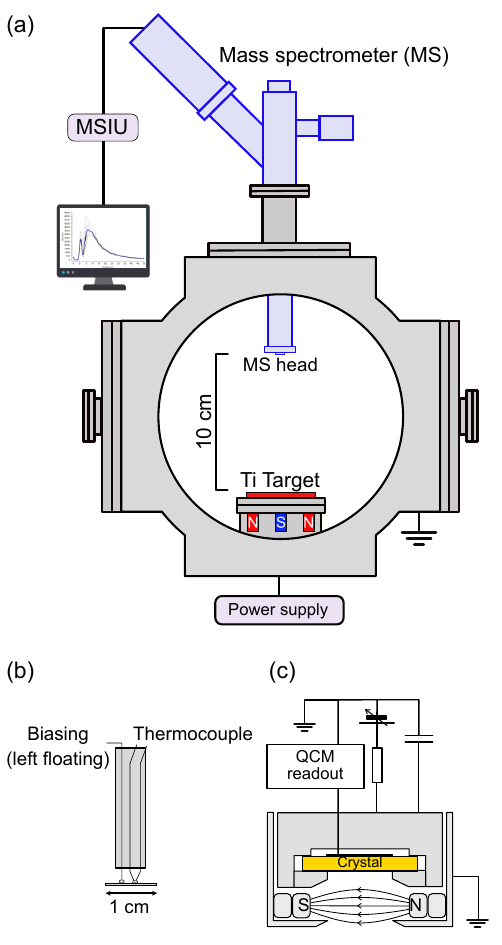}
    \caption{(a) Schematic of the experimental setup with a mass spectrometer (MS). (b) Passive thermal probe with an integrated thermocouple used for temperature measurements and a separate wire for external bias connection (left floating). (c) QCM ion meter with an external magnetic field, which prevents electrons from reaching the QCM crystal. The mass spectrometer head has been raised to accommodate either the thermal probe or the QCM ion meter at the same distance from the target.}
    \label{fig:EXP}
\end{figure}

The experiments were conducted in a stainless-steel vacuum chamber constructed from DN200ISO-K six-way cross tubing (Fig. \ref{fig:EXP}a). The chamber was equipped with a planar circular magnetron (VT100, Gencoa) with a tunable magnetic field and a 4-inch titanium (Ti) target. A turbo-molecular pump backed by a scroll pump maintained a base pressure of $4\times10^{-4}\,\mathrm{Pa}$. 

The magnetic field configuration used for all experiments can be denoted as C5E5, i.e. both the centre (C) and edge (E) magnets are moved 5 mm away from the target. The magnetic field is unbalanced, similar in shape to C0E0 (when both magnets are at positions closest to the target), but with a reduced strength, resulting in weaker plasma confinement above the racetrack than with C0E0. A representation of the magnetic field lines for similar configurations can be found in our earlier work \cite{Farahani2024R}.

\subsection{Pulse configurations}

The magnetron was operated in standard HiPIMS (s-HiPIMS) and chopped HiPIMS (c-HiPIMS) modes using a custom-built pulsing unit powered by a DC power supply (GS10, ADL) and controlled by a waveform generator (Rigol DG4102). Argon (Ar) was used as the working gas at a pressure of 1 Pa, with a constant flow rate of 40 sccm. 
The target current was monitored using a current probe (TCP303, Tektronix), and the target voltage was measured using a high-voltage probe (TESTEC, 100× attenuation). All signals were recorded using a digital oscilloscope (Picoscope 5444D, Pico Technology), which enabled precise power measurements.

The average power delivered during the pulses (average pulse power) was maintained at 40 kW. The s-HiPIMS pulse is defined by the total pulse length and repetition frequency. The reference s-HiPIMS was operated with $130\,\mathrm{\mu s}$ pulse length and $96\,\mathrm{Hz}$ frequency. The c-HiPIMS pulse configuration is defined using three parameters and denoted as $5$($S_{\mathrm{1}}$, $S_{\mathrm{o}}$, $D$), where 5 is the number of segments (fixed in this work), $S_{\mathrm{1}}$ is the length of the first pulse segment, $S_{\mathrm{o}}$ is the length of the other four segments, and $D$ is the delay time between segments, resulting in a total of five sequential pulses per cycle. The first micropulse is intentionally chosen longer to assist with plasma initiation, while the subsequent micropulses are shorter. The average power was always 500 W. When the total pulse length was varied, the frequency was adjusted accordingly to maintain both the pulse and the average power at the specified values. 

\subsection{Substrate flux characterization}

To estimate the total energy flux from the plasma to the substrate, a passive thermal probe, operated in floating mode, was placed 100 mm in front of the target (Fig. \ref{fig:EXP}b). The energy input to the probe was derived from the rate of temperature change during plasma-on and plasma-off phases according to: 

\begin{equation}
Q_\mathrm{in} = C_\mathrm{p} \left[ \left( \ddt{\Tp} \right)_\mathrm{heat} - \left( \ddt{\Tp} \right)_\mathrm{cool} \right],
\label{eq:Qin}
\end{equation}

where $Q_{\mathrm{in}}$ is the energy flux transferred to the probe, $C_p$ is the heat capacity of the thermal probe, $\left( \frac{dT_p}{dt} \right)_{\mathrm{heat}}$ is the rate of temperature increase during the plasma-on phase, and $\left( \frac{dT_p}{dt} \right)_{\mathrm{cool}}$ is the rate of temperature decrease during the cooling phase after plasma shut-off. It is important to note that the energy flux comprises contributions from all plasma species, including Ar atoms and ions, Ti atoms and ions, and electrons. A detailed description of the thermal probe design, operation, and post-processing methodology can be found in \cite{Farahani2024}.

A quartz crystal microbalance (QCM) ion meter was employed to measure the mass flux of deposited species (Fig. \ref{fig:EXP}c), namely Ti atoms and ions. It is based on the standard front-loaded QCM sensor (Inficon), positioned 100 mm from the Ti target and facing the target centre. To suppress electrons reaching the front of the QCM and the crystal, a pair of cylindrical SmCo magnets (10 mm in length and 8 mm in diameter) was mounted beneath the QCM, forming a localised magnetic field parallel to the crystal surface that suppresses electron flux to the sensor. These magnets were fixed to a grounded stand, and the distance between the magnets (15 mm) was chosen to minimise restriction on species reaching the QCM crystal. The QCM and magnets were also shielded from the front and sides by a grounded cylindrical cover with a 10 mm diameter aperture to ensure that atoms and ions are sampled from the front as if they were impacting on a flat grounded substrate. The body of the QCM was wrapped with Kapton tape to isolate it from the ground and limit electron currents. Note that the diameter of the opening directly in front of the crystal was 8 mm. A DC bias power supply (HCS-3104, Manson) was connected to the QCM chassis to repel ions with a positive bias voltage. A resistor (16 $\Omega$) is added in series with the power supply to protect it from high currents, and a capacitor (470 $\mu$F) is placed in parallel to stabilize the bias voltage when high current is drawn from the plasma.

When the DC bias voltage is set to $0\,\mathrm{V}$, the total deposition rate $R_\mathrm{t}$, including Ti atoms and ions, is measured by the QCM. Then, a positive bias of $25\,\mathrm{V}$ is set to repel positive ions, enabling selective measurement of neutral Ti deposition rate, $R_\mathrm{n}$. The ion contribution to the deposition rate, $R_\mathrm{i}$, is then determined as
\begin{equation}
R_\mathrm{i} = R_\mathrm{t} - R_\mathrm{n} \,.
\end{equation}
The deposition rates were obtained from the mass gain accumulated over 5 minutes.

In situ mass- and energy-resolved ion flux measurements were carried out using a mass spectrometer (MS, EQP300, Hiden Analytical). The detected ionic species include \ioni{36}{Ar}{1}, \ioni{50}{Ti}{1}, \ioni{40}{Ar}{2}, and \ioni{48}{Ti}{2}. Due to saturation of the secondary electron multiplier (SEM) detector when measuring the more abundant isotopes (\ioni{40}{Ar}{1} and \ioni{48}{Ti}{1}), the less abundant isotopes \ioni{36}{Ar}{1} and \ioni{50}{Ti}{1} were used to represent singly-charged argon and titanium ions, respectively. All recorded counts were subsequently corrected using the corresponding natural isotopic abundance factors. Energy distribution functions were integrated over the energy range 0–20 eV to determine the total ion fluxes reaching the detector. The MS head was lifted to allow space for connecting either the passive thermal probe or the QCM ion meter. During mass spectrometry measurements, the probes were removed, and the spectrometer head was lowered so that its orifice aligned with the probes' position, 100 mm from the target centre.

\section{Results}

\subsection{Pulse configuration waveforms}

\begin{figure} 
    \centering
    \includegraphics[width=85mm]{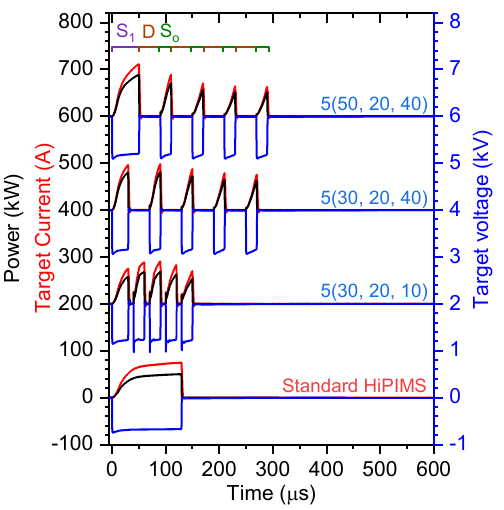}
    \caption{Selected waveforms of the target voltage, the target current, and the power for s-HiPIMS and various c-HiPIMS pulse configurations denoted as 5($S_{\mathrm{1}}$, $S_{\mathrm{o}}$, $D$), where $S_{\mathrm{1}}$ is the length of first pulse, $S_{\mathrm{o}}$ is the length of other pulses and $D$ is the delay between pulses. The average power and pulse power were held constant across all conditions.}
    \label{fig:waveform}
\end{figure}

Figure \ref{fig:waveform} presents the temporal evolution of target voltage, target current, and discharge power for a s-HiPIMS discharge and several c-HiPIMS configurations under the weaker unbalanced magnetic field configuration (C5E5). 

In the s-HiPIMS configuration, the voltage waveform is nearly rectangular, showing a small drop after the initial rise before stabilising. The corresponding current waveform exhibits an initial rise with a little delay caused by the time required to initiate the discharge from a low initial density of residual charge carriers \cite{Hnilica2023, Anders2009, Yushkov2010}, before reaching a steady plateau that reflects sustained plasma conditions. 

In the 5(30, 20, 10) configuration, the first micropulse exhibits an initial delay in current onset and reaches a lower peak current compared to the second and third segments. This behaviour is commonly observed and is attributed to the relatively low density of residual charge carriers, such as electrons, argon ions, and metal ions, present at the start of the pulse sequence \cite{Hnilica2023, Soucek2021, Barker2013}. In contrast, the second and third micropulses exhibit a steeper current rise and higher peak values, due to residual charge carriers remaining in the near-cathode region during the brief 10 $\mu$s delays. These short intervals preserve ion and electron populations, allowing for faster discharge ignition and higher current. As the sequence progresses, cumulative effects such as gas heating and sputtering wind lead to a depletion of the working gas near the target (gas rarefaction), reducing the ionisation probability and, consequently, the discharge current \cite{Hnilica2023, Barker2013, Kubart2011}. 

In the second c-HiPIMS configuration, 5(30, 20, 40), the pulse sequence is similar to the previous configuration, but with a longer 40 $\mu$s delay between pulses. As in the earlier case, the first micropulse exhibits a delayed current onset; however, its peak current is comparable to that of the second micropulse and higher than for the 5(30, 20, 10) case, which can be explained by the slightly higher voltage necessary to sustain the same pulse power. The subsequent pulses continue to exhibit faster current rise rates due to residual charge carriers remaining from prior discharges. Despite this, the overall peak currents in the later pulses are slightly reduced and eventually stabilise, indicating a steady-state discharge condition under the extended off-time. 

In the final c-HiPIMS configuration, 5(50, 20, 40), the first micropulse is longer (50 $\mu$s) than in previous modes and exhibits a markedly higher peak current compared to both the subsequent pulses in the same sequence and the pulses observed in earlier configurations. The elevated peak is attributed to the extended length of the first pulse, which allows the discharge to fully evolve. The following 20~$\mu$s micropulses, although characterised by a faster current rise due to residual charge carriers, do not reach the same peak current as the initial pulse. The peak current progressively decreases across the later pulses until stabilising (in the last two pulses). This decrease is influenced not only by reduced charge carrier availability and possible gas rarefaction, but also by power balancing, since the pulse power and average power are fixed, the high initial peak results in a high power input early in the sequence, limiting the allowable power, and thus the peak current, of the later segments to remain within the system's average power constraints. 

\subsection{Pulse length in chopped HiPIMS}

\begin{figure} 
    \centering
    \includegraphics[width=85mm]{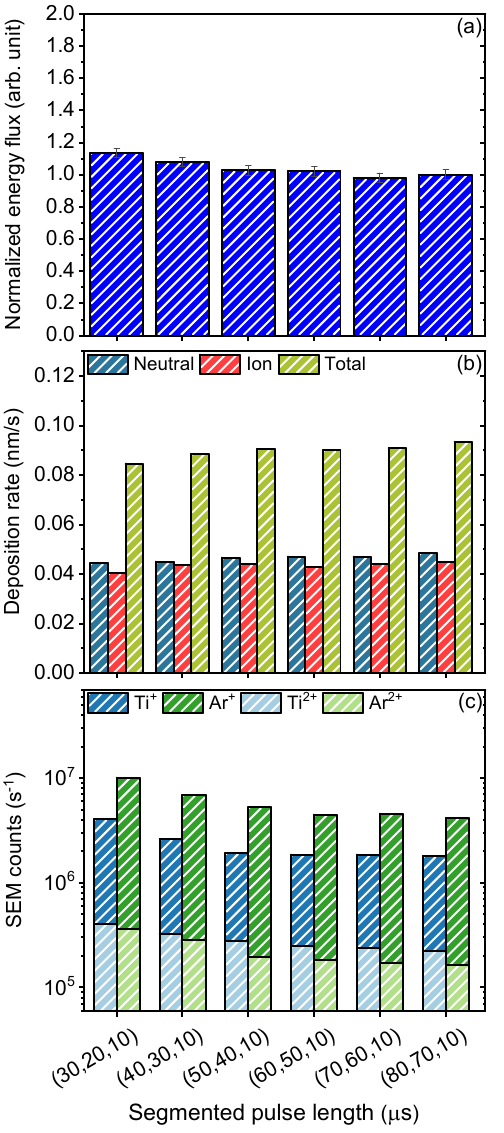}
    \caption{(a) Energy flux measured by the thermal probe at floating potential, (b) neutral Ti atom, Ti ion, and total deposition rates measured by the QCM ion meter, and (c) integrated ion energy flux distribution functions over the energy range 0–20~eV measured by MS, for various chopped HiPIMS pulse configurations. The first and other pulse lengths ($S_{\mathrm{1}}$, $S_{\mathrm{o}}$) were increased by $10\,\mu\mathrm{s}$ in each step. The average power and pulse power were held constant across all conditions.}
    \label{fig:pulsetime}
\end{figure}

Figure \ref{fig:pulsetime} shows the effect of the pulse length on the energy flux, the deposition rate, and Ti and Ar ion fluxes reaching the substrate at a distance of 100 mm from the target centre. The lengths of the first, $S_{\mathrm{1}}$, and the other pulses, $S_{\mathrm{o}}$, were increased incrementally by 10 $\mu$s at each step, while the delay time was held constant at 10 $\mu$s. Let us remind you that the frequency was changed to maintain the same pulse and average power. The energy flux is normalised to the 5(80, 70, 10) pulse configuration. 

As shown in \ref{fig:pulsetime}a, when the pulse lengths increase, the energy flux very slightly gradually decreases. Similarly, the MS results also revealed a decrease in the intensity of both singly and doubly ionised species as the pulse length increased from 5(30, 20, 10) to 5(60, 50, 10), after which the ion signals remained constant, see Figure~\ref{fig:pulsetime}c. Notably, the difference in signal intensity between \ion{Ti}{2} and \ion{Ar}{2} became more pronounced with longer pulse segments. 

In contrast to \ref{fig:pulsetime}a and \ref{fig:pulsetime}c, Figure \ref{fig:pulsetime}b shows a very slight increase in the deposition rate as the pulse length increases, while the IFF remains constant. Overall, the observed changes are small, especially compared to the other trends reported in this paper, so we conclude that the pulse length has no significant effect on the deposition rate, IFF, or energy flux under the investigated conditions.

\subsection{Pulse delay in chopped HiPIMS}

\begin{figure} 
    \centering
    \includegraphics[width=85mm]{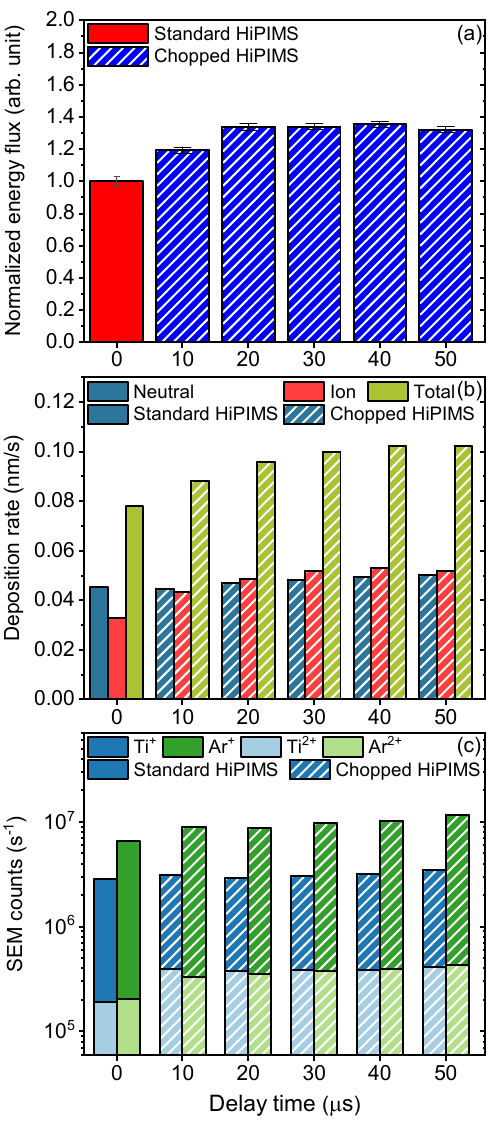}
    \caption{(a) Energy flux measured by the thermal probe at floating potential, (b) neutral Ti atom, Ti ion, and total deposition rates obtained from the QCM ion meter, and (c) integrated ion energy flux distribution functions over the energy range 0–20~eV measured by MS, for various pulse configurations, standard ($130\,\mu\mathrm{s}$) and chopped HiPIMS 5(50, 20, \textit{D}). The delay time between segments (\textit{D}) was increased by $10\,\mu\mathrm{s}$ in each step. The average power and pulse power were held constant across all conditions.}
    \label{fig:delay}
\end{figure}

Figure \ref{fig:delay} illustrates the effect of the delay time between individual pulses of the c-HiPIMS configurations on the energy flux, the deposition rate, and Ti and Ar ion fluxes at the substrate for the C5E5 magnetic field configuration. For the c-HiPIMS, the pulse configuration 5(50, 20, \textit{D}) was used, where only the delay time (\textit{D}) between micropulses was varied in 10 $\mu$s increments, starting from 10 $\mu$s. The delay of 0 $\mu$s corresponds to the s-HiPIMS. In all cases, the total length of the negative pulses was fixed at 130 $\mu$s. The energy flux is normalised to the s-HiPIMS case. 

Overall, c-HiPIMS exhibits a higher energy flux compared to s-HiPIMS (Fig. \ref{fig:delay}a). In the chopped mode, the lowest energy flux is observed at the shortest delay time (10 $\mu$s). As the delay time increases, the energy flux initially increases and then saturates at a constant value.

Figure \ref{fig:delay}b shows that the total deposition rate also gradually increases and tends to a saturation with increasing delay time. We observe only a minor increase in the neutral-atom flux, while the increase in ion flux is clearly the dominant component. In other words, the IFF clearly increases with a shift to c-HiPIMS and with increasing the pulse-to-pulse delay. This supports the increase in the energy flux shown in Figure \ref{fig:delay}a.

As shown in Figure \ref{fig:delay}c from the MS data, s-HiPIMS exhibits lower ion intensities, particularly for \ion{Ar}{1} and doubly ionised species. When the HiPIMS pulse is chopped, increasing the delay slightly increases ion intensities, consistent with thermal probe and ion meter data. The increase is slightly more pronounced for Ar ions. Notably, the fraction of doubly charged ions is higher for c-HiPIMS configurations compared to the s-HiPIMS configuration.

\subsection{Pulse length in standard HiPIMS }

The previous section described the effect of the pulse delay in the c-HiPIMS mode. If the pulse delay in the five-segment c-HiPIMS is increased even more (beyond the range presented in the previous section), it will converge on a regime similar to s-HiPIMS with 5 times shorter pulse length and 5 times higher frequency. Therefore, to facilitate a meaningful discussion of the benefits of c-HiPIMS, we also investigated s-HiPIMS discharges with varying pulse lengths and repetition frequencies, while keeping the average and pulse power constant. For a reference pulse length of 130 $\mu$s, the repetition frequency was 96~Hz. For other pulse lengths, $t_\mathrm{on}$, the repetition frequency scales as $\propto 1/t_\mathrm{on}$.

\begin{figure}
    \centering
    \includegraphics[width=85mm]{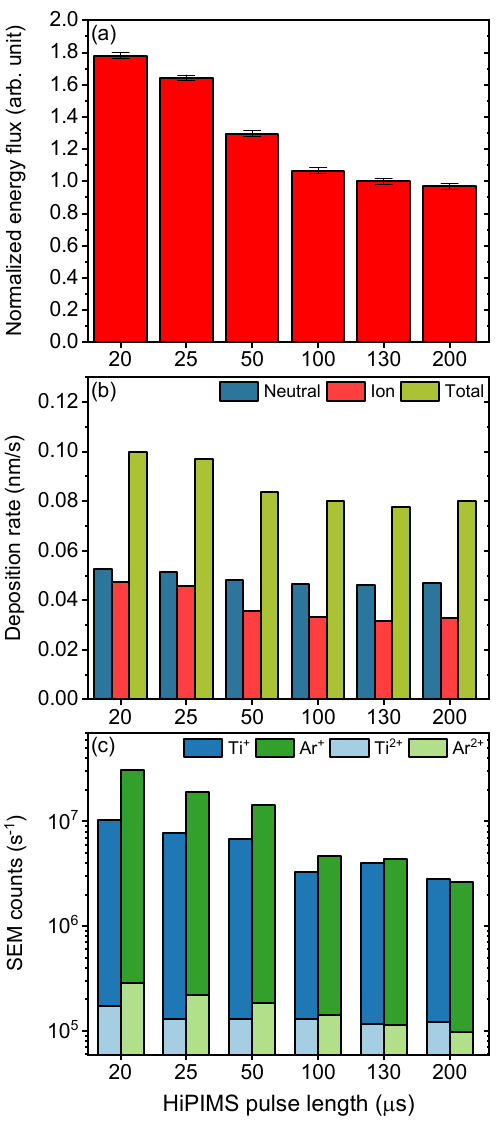}
    \caption{Energy flux measured by the thermal probe at floating potential, (b) neutral Ti atom, Ti ion, and total deposition rates obtained from the QCM ion meter, and (c) integrated ion energy flux distribution functions over the energy range 0–20~eV measured by MS, for HiPIMS discharges with varying pulse lengths and repetition frequencies at constant average and pulse power. The average power and pulse power were held constant across all conditions.}
    \label{fig:STHiPIMS}
\end{figure}

Figure \ref{fig:STHiPIMS} presents the effects of these variations on energy flux, deposition rate, and the fluxes of Ti and Ar ions at the substrate, measured under the C5E5 magnetic field configuration. The energy flux is normalised to the 130 $\mu$s s-HiPIMS configuration. 

The thermal probe measurements, as shown in Figure \ref{fig:STHiPIMS}a, reveal a pronounced decrease in energy flux as the pulse length increases from 20 to 100 $\mu$s. This initial drop is followed by a continued, but considerably slower, decline extending to the 200 $\mu$s pulse length.

A comparable trend is observed in the QCM ion meter data (Figure \ref{fig:STHiPIMS}b). The deposition rate decreases more sharply between 20 and 100 $\mu$s, then saturates. The Ti atom contribution decreases only very slightly, but the Ti ion contribution exhibits a sharp decrease between 20 and 50 $\mu$s. Beyond 100 $\mu$s, both atom and ion contributions are practically constant.

The MS results also follow this general pattern (Figure \ref{fig:STHiPIMS}c). Ion counts decrease substantially between 20 and 100 $\mu$s, then decline more slowly up to 200 $\mu$s. For pulse lengths of 100 $\mu$s or longer, we observe comparable ion counts for Ar and Ti ions (both singly and doubly charged). On the other hand, for pulse lengths of 50 $\mu$s and shorter, the Ar ion species counts increase more rapidly and, as a result, the \ion{Ar}{1}/\ion{Ti}{1} ratio increases as the pulse length decreases.

\section{Discussion}

\subsection{Diagnostic methods}

This section highlights the specifics of each diagnostic method used. The thermal probe responds to the total energy flux from all incoming species, including Ti atoms and ions, Ar atoms and ions, and electrons. Since the probe is at a floating potential, it is expected that the electron current equals the ion current and that ions contribute more energy per arriving particle than neutral atoms. The QCM ion meter, in contrast, measures only the absolute particle flux of film-forming species, is sensitive to Ti atoms and ions, and reflects incoming Ar. A key advantage of the QCM ion meter and the thermal probe is their large acceptance area. They are not sensitive to the particle's angle of incidence (except for the aperture effect in the case of the ion meter, where the particles entering the aperture must still travel about 10 mm past the shielding magnets to the crystal surface). On the other hand, the MS measures the particle flux of only charged species, but it discerns between Ar and Ti species and their charge states. The SEM counts cannot be easily converted to particle flux densities. MS samples ions through a small orifice and within a limited acceptance angle, making it less sensitive to ions arriving from off-axis directions. The overall sensitivity depends on the ion energy and the ion species' mass and can vary with plasma conditions due to the considerable sensitivity to the specific tuning of the MS. Overall, the results in Figures \ref{fig:pulsetime}, \ref{fig:delay}, and \ref{fig:STHiPIMS} show very good agreement among the three diagnostic techniques, so we can benefit from their complementarity in the discussion.

A small discrepancy is observed in  Figure~\ref{fig:pulsetime} between the trends obtained by the thermal probe and MS and the trend obtained by the QCM ion meter. In this case, it should be noted that earlier measurements using a different geometry of the QCM ion meter also showed a slight decrease in deposition rate with increasing pulse length, consistent with the thermal probe and MS results (but with excessively high absolute IFF values). The presented results were remeasured some time after the corresponding thermal probe measurements, so the slight discrepancy can be attributed to target erosion and the resulting changes in discharge I-V characteristics and radial transport. Therefore, we deduce that other process parameters (target erosion, magnetic field configuration, etc.) might have a comparably strong effect on the measured quantities as the pulse length does.

\subsection{HiPIMS plasma phenomena}

To understand the presented results, it is necessary to consider several mechanisms that govern the discharge dynamics and the transport of target material species to the substrate.

At the start of a pulse, sputtering is efficiently driven by process gas ions generated from the background gas. However, as the target material atoms are sputtered from the target, they become ionised (depending on their ionisation cross section) and these ions are drawn back to the negatively biased target surface by the electric field formed in the magnetic trap region (magnetic presheath). The potential drop across this region is on the order of tens of volts \cite{Mishra2011a, Rauch2012, Panjan2017, Held2022}. This ion back-attraction is responsible for the decrease in deposition rate compared to sputtering driven by only process-gas ions \cite{Anders2011, Huo2014, Anders2008, Gudmundsson2012}. Lower ion back-attraction means more target-material ions can leave the magnetic trap region and be deposited on the substrate, thereby increasing the IFF and deposition rate \cite{Vlcek2010}.

Moreover, the spatial dynamics of ion transport can contribute to the loss of target material en route to the substrate, particularly via radial transport \cite{Bohlmark2006, Lundin2008}. In HiPIMS, acceleration of ions tangentially outward from the racetrack and the associated increased radial material loss have been reported \cite{Lundin2008, Poolcharuansin2011}. This was explained by azimuthal electric-field oscillations associated with the rotating ionisation zones (spokes) \cite{Panjan2017, Held2020}. Although not exclusively dependent on pulse length, longer pulses provide greater temporal windows in which such radial transport mechanisms can develop. Velicu et al. \cite{Velicu2017} showed that adding an external magnetic field can limit the radial ion losses, increase deposition rate, ionised flux fraction, and ion energy for a tungsten target. 

Gas rarefaction in front of the target occurs due to the momentum transfer from energetic sputtered atoms to the background process gas atoms, leading to the displacement of Ar away from the target \cite{Hoffman1985, Rossnagel1988}. Naturally, this effect becomes stronger with increasing current, pulse length and sputtering yield \cite{Barynova2024}. This local depletion of working gas limits further \ion{Ar}{1} production and reinforces dependence on the self-sputtering process. A higher density of the process gas supports a rapid rise in discharge current at the beginning of the pulse due to efficient secondary electron emission.  The gas depletion during the pulse naturally leads to a greater contribution from metal-ion sputtering, which negatively impacts the deposition rate, as discussed above.

Although we understand the role of these phenomena, the subtle differences in measured plasma and deposition characteristics arise from their complex interplay and balance, which are difficult to predict.

\subsection{Effect of pulse length and pulse chopping}

Several earlier studies have shown that using shorter pulses, either in one-pulse or multi-pulse mode, is beneficial for achieving higher deposition rate and ionised flux fraction. This was attributed to reduced ion back-attraction \cite{Lundin2012, Tiron2015, Rudolph2020}, reducing gas rarefaction, and suppressing the onset of self-sputtering \cite{Huo2021, Barker2013, Tiron2018, Antonin2015, Tiron2018a}. Shimizu et al. \cite{Shimizu2021} reported that, at constant average power and peak current, the deposition rate increases with shorter pulse lengths, reaching a maximum at about 25–50 $\mu$s. Similarly, Barker et al.\cite{Barker2013} found that at a constant average power, c-HiPIMS with four 25 $\mu$s micropulses achieved a higher deposition rate than with two 50 $\mu$s micropulses. The most important factor seems to be that, after a short pulse, when the target material is sputtered and ionised, the pulse is terminated, weakening the electric field in the magnetic trap region and allowing the ions to escape toward the substrate (c.f. the reduced ion back-attraction).

However, the above-mentioned studies did not always maintain the peak current or pulse-averaged power constant to ensure a comparable degree of ionisation, nor did some evaluate the IFF alongside the deposition rate, which makes it difficult to draw definitive conclusions.

The present systematic investigation definitely confirms this trend and provides further insight. First, s-HiPIMS with a shorter pulse length and the same pulse power density clearly shows increased deposition rate, energy flux, and IFF; see Figure~\ref{fig:STHiPIMS}. Dividing a HiPIMS pulse into multiple shorter segments (i.e., chopping the pulse) yields a similar outcome -- see the significant boost in the IFF between the standard and any c-HiPIMS configuration in Figure~\ref{fig:delay}. Finally, even among the chopped configurations, making the pulses shorter yields a higher energy flux, though the effect is more subtle here; see Figure~\ref{fig:pulsetime}. It should be noted that in ref.~\cite{Hnilica2025}, Hnilica et al. reported a slight increase in deposition rate and a decrease in IFF when the number of pulses was increased from one (s-HiPIMS) to up to 32 (c-HiPIMS) at a fixed pulse power density (series C). Only for 2 micropulses did the IFF increase. This is in partial contradiction with our results, which show an increase in both the deposition rate and IFF. We speculate that these differences can be due to differences in the experimental setup, such as the magnetron and magnetic field design.

Most importantly, our results systematically demonstrate that the delay between chopped pulses is a critical parameter for maximising both deposition rate and energy flux, with an optimal off-time of at least 30 $\mu$s (Fig. \ref{fig:delay}). The introduction of longer $D$ serves two main purposes. First, it provides a sufficient window for sputtered metal ions to escape the near-target region and reach the substrate, preventing them from being back-attracted upon the following pulse's reignition. Second, the increased \ion{Ar}{1} signal at longer delays indicates that this off-time allows for partial replenishment of the working gas. This creates more favourable conditions for the initiation of the next micropulse, promoting efficient initial sputtering via \ion{Ar}{1} bombardment, but at the same time, target material species are still abundant and available to be quickly ionised\cite{Hnilica2023, Hnilica2026}.

The results of Barker et al.\cite{Barker2013, Barker2014} are consistent with our study, showing a higher deposition rate for multiple short pulses with long off-times than for s-HiPIMS at the same total negative pulse length. In their work, however, the deposition rate saturated for longer pulse delays ($\sim 200\,\mu\mathrm{s}$). Antonin et al. \cite{Antonin2015} varied the delay between micropulses for a tungsten target at constant average power with pre-ionisation. They found an optimal delay of 50 $\mu$s, which maximised the deposition rate and corresponded to peaks in both ion flux and neutral density. Recently, Hnilica et al. \cite{Hnilica2026} studied pulse configurations very similar to ours, with 3 pulses and variable delays between them. They observed a similar, yet smaller, increase in deposition rate when the delay increased from 10 to 120 $\mu$s. On the other hand, they observed a small decrease in the ionised flux fraction, which again partially contradicts our results. Again, this might be  due to differences in the magnetron and magnetic field design.

\subsection{Comparing standard and chopped HiPIMS}

\begin{figure}
    \centering
    \includegraphics[width=130mm]{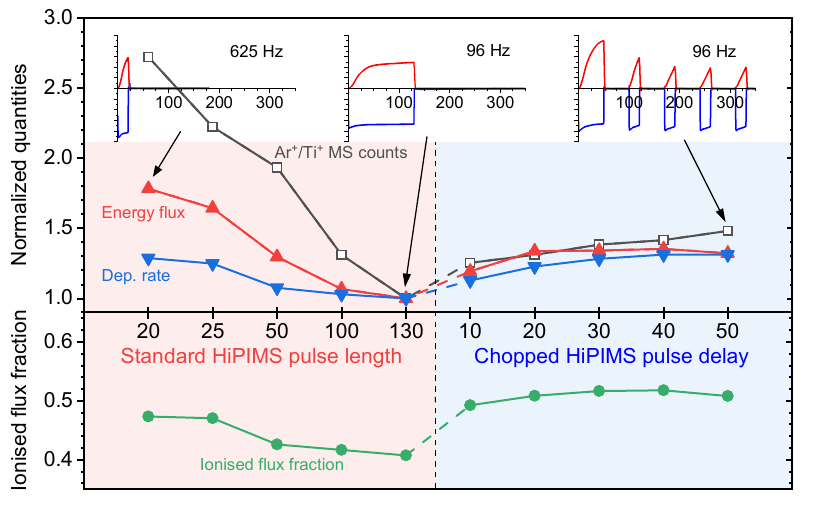}
    \caption{Direct comparison of measured quantities for s-HiPIMS with a variable pulse length and repetition frequency (left side, red background) and the chopped HiPIMS configurations 5(50,20,$D$) with a variable pulse delay $D$ (right side, blue background). Deposition rate, energy flux, IFF and \ion{Ar}{1}/\ion{Ti}{1} ion count ratio, all normalised to the reference case (s-HiPIMS with 130 $\mu$s pulse length), are shown. The average power and average pulse power are fixed for all configurations. The insets at the top show the voltage and current waveforms for selected configurations.}
    \label{fig:comparison}
\end{figure}

We have shown that both chopping a long HiPIMS pulse into multiple shorter pulses and shortening the single pulse and increasing the repetition frequency are beneficial for increasing deposition rate and IFF. It remains to be discussed whether there are any potential differences between the two approaches. 

The main distinction between c-HiPIMS and s-HiPIMS at short pulse lengths and high frequencies lies in the delay time introduced between micropulses in the chopped mode. For instance, by progressively increasing the delay in the 5(50, 20, D) configuration, the system closely approximates the operating conditions of s-HiPIMS with a 20 $\mu$s pulse length at a five times higher frequency. We have verified that the length of the first micropulse does not play a crucial role, as the differences in energy flux and IFF between the 5(50,20,40) and 5(20,20,40) configurations are insignificant (not shown). 

A compilation of the results presented in Figs.~\ref{fig:delay} and \ref{fig:STHiPIMS} for direct comparison is shown in Fig.~\ref{fig:comparison}. It shows all relevant quantities, i.e. the deposition rate, energy flux, IFF, and the \ion{Ar}{1}/\ion{Ti}{1} ion count ratio normalised to the reference case, which is the s-HiPIMS with 130 $\mu$s pulse length (in the middle of the graph). It is emphasised that both the average power and the pulse power were held constant, ensuring comparable degree of ionisation. To the left of the reference case, the s-HiPIMS pulse length decreases (red background). To the right of the reference case, the delay $D$ between the five micropulses in the 5(50,20,$D$) configurations increases (blue background). This figure again shows that all watched quantities increase relative to the reference case as the pulse length is shortened and the pulse delay is increased.

Eventually, we can focus on the best configuration from each regime, i.e. s-HiPIMS with a 20 $\mu$s pulse length operated at the highest repetition frequency of $625\,\mathrm{Hz}$ and the chopped configuration with the highest delay, 5(50, 20, 50) operated at $96\,\mathrm{Hz}$. The same deposition rate was measured in both cases, while the IFF was slightly higher in the c-HiPIMS case. On the other hand, the energy flux was significantly higher for the s-HiPIMS case. It should be noted that quantitative comparisons of results from different series of experiments conducted at different times may be subject to greater systematic error due to changing plasma conditions (target erosion) and thermal probe capacitance (material deposited on the probe). A repeated measurement of the energy flux for both extremal configurations, performed in immediate succession, confirmed the higher energy flux for the s-HiPIMS case, but the difference was only $11\,\%$ (not shown). 

The higher energy flux may seem to be in contradiction with the comparable deposition rate and IFF in both cases. However, it can be explained by the more prominent increase of the \ion{Ar}{1}/\ion{Ti}{1} ion count ratio (from the reference case) in s-HIPIMS (to the left) compared to c-HiPIMS (to the right), which indicates that the ion flux composition contains a higher fraction of \ion{Ar}{1} ions in the s-HiPIMS configurations. The long off-time in s-HiPIMS (1580 $\mu$s) allows near-complete refilling of Ar near the target compared to the short ($<50\,\mu$s) off-time in c-HiPIMS. Note that a complete Ar refill requires off-times on the order of a few ms~\cite{Lundin2009, Kozak2023}. Therefore, the higher energy flux in the case of s-HiPIMS can be attributed to a higher fraction of process-gas ions in the flux to the substrate. We conclude that both s-HiPIMS and c-HiPIMS configurations, when optimised, can achieve similar performance in deposition rate and IFF, but c-HiPIMS is more effective at suppressing the potentially unwanted flux of process-gas ions.

\section{Conclusions}

Chopped HiPIMS was systematically investigated under constant average power and constant pulse-averaged power to elucidate its effect on the deposition rate, ionised flux fraction, and the energy delivered to the substrate, which are the main parameters that strongly influence film growth. The study combined the results of total energy flux, fluxes of titanium atoms and ions to the substrate, and ion counts from the mass spectrometer to provide a detailed understanding of the deposition conditions. All complementary measurements were in very good agreement.

It was shown that chopped HiPIMS generally improved energy flux and deposition rates compared to standard HiPIMS with the same total negative pulse length. Among the chopped HiPIMS configurations, the measured quantities showed only minor changes as the length of individual micropulses was varied. On the other hand, increasing the delay time between micropulses significantly enhanced the energy flux, deposition rate and ionised flux fraction, with an optimal off-time of at least 30 $\mu$s.

Finally, chopped HiPIMS configurations were compared to standard HiPIMS operated at higher frequencies with consistently short pulse lengths. The standard HiPIMS regime with a short 20 $\mu$s pulse yielded higher energy fluxes, but the same deposition rate and slightly lower ionised flux fraction compared to the chopped HiPIMS configuration with a pulse delay of $50\,\mu$s. The higher energy flux can be attributed to the increased fraction of process gas (Ar) ions.

We conclude that an optimised application of short pulses, even in the standard HiPIMS configuration, is beneficial for increasing the deposition rate, the ionised flux fraction, and the total energy flux to the substrate. This is consistent with previous findings, which highlight the effect of ion release from the target vicinity during off-times. Moreover, we conclude that chopped HiPIMS with short pulses and off-times of around $40\mu$s between pulses can achieve similar or even slightly higher performance, with the added benefit of limiting the fluxes of process-gas ions.

\ack{This work was supported by the project Quantum materials for applications in sustainable technologies (QM4ST), funded as project No. CZ.02.01.01/00/22\_008/0004572 by Programme Johannes Amos Comenius, call Excellent Research.}





\providecommand{\newblock}{}

\end{document}